\documentclass[conference]{IEEEtran}
\IEEEoverridecommandlockouts
\usepackage{cite}
\usepackage{amsmath,amssymb,amsfonts}
\usepackage{algorithmic}
\usepackage{graphicx}
\usepackage{textcomp}
\usepackage{float}
\usepackage{subfig}
\usepackage{color}
\usepackage[pagebackref,breaklinks,colorlinks]{hyperref}
\usepackage[capitalize]{cleveref}
\usepackage{booktabs}
\usepackage[table]{xcolor}
\usepackage{bm}
\usepackage{multirow}
\newcommand{\tablestyle}[2]{\setlength{\tabcolsep}{#1}\renewcommand{\arraystretch}{#2}\centering\footnotesize}
\def\BibTeX{{\rm B\kern-.05em{\sc i\kern-.025em b}\kern-.08em
    T\kern-.1667em\lower.7ex\hbox{E}\kern-.125emX}}
\begin{document}

\title{Assessing and Enhancing Robustness of Deep Learning Models with Corruption Emulation in Digital Pathology\\
}

\author{
    \IEEEauthorblockN{
    Peixiang Huang$^{1*}$, Songtao Zhang$^{1*}$, Yulu Gan$^1$, Rui Xu$^1$, Rongqi Zhu$^1$, Wenkang Qin$^1$,\\Limei Guo$^2$, Shan Jiang$^3$, Lin Luo$^{1\dagger}$
    }
    \IEEEauthorblockA{$^1$ College of Engineering, Peking University, Beijing, China}
    \IEEEauthorblockA{$^2$ Third Hospital, Peking University Health Science Center, Beijing, China}
   \IEEEauthorblockA{$^3$ Institute of Biomedical Engineering, Beijing Institute of Collaborative Innovation, Beijing, China}
    \IEEEauthorblockA{\{huangpx, ganyulu, xurui, qinwk\}@stu.pku.edu.cn, \{songtzhang, luol\}@pku.edu.cn, \\ rongqizhu77@gmail.com, guolimei@bjmu.edu.cn, jiangs@jingjinji.cn \
    } 
    \thanks{$*$ These authors contributed equally to this work.}
    \thanks{$\dagger$ Corresponding author.}
}

\maketitle

\begin{abstract}
Deep learning in digital pathology brings intelligence and automation as substantial enhancements to pathological analysis, the gold standard of clinical diagnosis. However, multiple steps from tissue preparation to slide imaging introduce various image corruptions, making it difficult for deep
neural network (DNN) models to achieve stable diagnostic results for clinical use. In order to assess and further enhance the robustness of the models, we analyze the physical causes of the full-stack corruptions throughout the pathological life-cycle and propose an Omni-Corruption Emulation (OmniCE) method to reproduce 21 types of corruptions quantified with 5-level severity. We then construct three OmniCE-corrupted benchmark datasets at both patch level and slide level and assess the robustness of popular DNNs in classification and segmentation tasks. Further, we explore to use the OmniCE-corrupted datasets as augmentation data for training and experiments to verify that the generalization ability of the models has been significantly enhanced.
\end{abstract}


\section{Introduction}
\begin{figure*}[!t]
    \centering
    \includegraphics[width=0.99\textwidth]{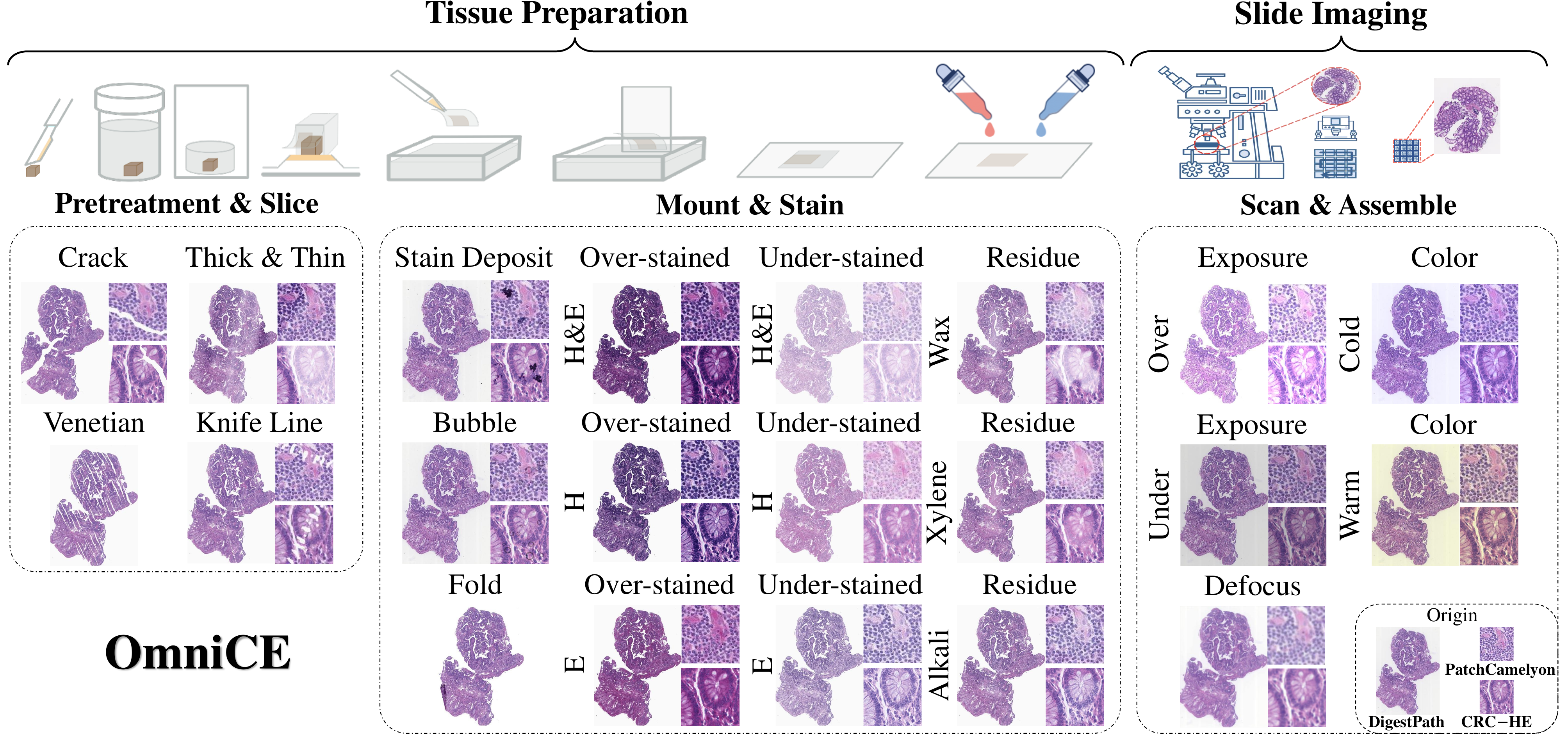}
    \caption{\textbf{OmniCE} corruptions on three benchmark datasets at the patch and slide level are generated based on the investigation of corruptions produced throughout the full pathological life-cycle. Two main phases, \textbf{tissue preparation} and \textbf{slide imaging}, consist of several substeps: \textbf{pretreatment \& slice} (extraction, fixation, dehydration, clearing, wax dipping, embedding, sectioning), \textbf{mount \& stain} (mounting, baking, dewaxing, H\&E staining, sealing), \textbf{scan \& assemble} (parameter setting, scanning, focusing, shooting, image stitching, compression). }
    \label{fig:taskflow}
\end{figure*}

Pathological diagnosis is the gold standard for precise diagnosis and treatment for most diseases especially tumors and cancers. 
Digital pathology with high-resolution scanned images of pathological slides enables the use of deep learning algorithms in helping pathologists improve diagnostic efficiency and quality \cite{qin2022pathtr}. 
However, the whole process of producing pathological slides and digital images (Fig. \ref{fig:taskflow}) involves various corruptions such as artifacts in specimen preparation and image processing. These corruptions challenge deep neural network (DNN) models in the diagnostic reliability under clinical circumstances.
DNN models are sensitive to general image corruptions such as Gaussian noises and Exposure variance and suffer from severe performance drop. The same problems occur on medical DNN with even worse impact.
To tackle the robustness issue of DNNs, several works have introduced adversarial data or corrupted samples to the training process. \cite{ChengXue2022RobustMI} trained a robust model for medical image classification from noisy-labelled data. \cite{BirgidSchmigMarkiefka2021QualityCS} conduct stress-testing on diagnostic models using synthetically generated artifacts for clinical validation.

In digital pathology, some works evaluate the performance of DNN models under different corruption types. \cite{YunlongZhang2022BenchmarkingTR} applies general image processing as corruptions on pathology images to benchmark robustness of various DNNs. \cite{BirgidSchmigMarkiefka2021QualityCS} digitally reproduces twelve types of pathological artifacts using both image processing and image style transfer methods and discovers performance loss of DNN models in prostate cancer detection. However, few method covers the full-stack pathological corruptions encountered along the digital pathology life-cycle, and most works focus on the visual similarities rather than the physical causes when reproducing the corruptions. Besides, how effective the model robustness can be improved against the corruptions are not explored in depth, 

In this paper we investigate how corruptions are generated throughout the full pathological life-cycle from tissue preparation to slide imaging. 21 types of corruptions are discovered, including \emph{Over/Under Stained with H}\&\emph{E/H/E}, \emph{Residual Wax/Xylene/Alkali}, \emph{Thick and Thin Section}, \emph{Over/Under Exposure}, \emph{Defocus}, \emph{Crack}, \emph{Venetian}, \emph{Fold}, \emph{Knife Line}, \emph{Bubble}, etc, while the causes such as human operations, materials, and device setups are analyzed. 

Accordingly, we propose an Omni-Corruption Emulation (OmniCE) method to simulate the physical mechanisms of the causes with mechanical engine, optical engine, chemical engine, etc, to reproduce the realistic and controllable image corruptions, with physicians' know-how in designing the scales. The OmniCE corruption benchmark datasets cover the 21 types of corruptions throughout the pathological life-cycle, each is quantified with the corruption severity of 5 levels, from shallow to deep. Three typical pathology datasets are applied with OmniCE and used to assess the influence on typical DNN models. 

Furthermore, we explore the use of the OmniCE corrupted data as augmentations of training data to improve DNN performance. We compared the datasets augmented by OmniCE vs by Augmix \cite{hendrycks2019augmix} which covers the common image processing types. Experimental results show that the model trained by OmniCE-augmentated datasets significantly outperforms that by Augmix-augmented ones of 8.3\% and 15.3\% on two different centers, and achieves the SOTA performance.

Our main contributions include: (1) We are the first to introduce the full-stack pathological corruption types that present throughout the digital pathology life-cycle. (2) We design the corruption emulator engines based on the underlying physical causes of corruptions to ensures that the corruptions reflect realistic clinical scenarios. (3) We experiment the pathology-specific corruption data as augmentation of training data and achieve a significant accuracy improvement in enhancing the model robustness compared with Augmix.

\begin{figure*}[!t]
	\centering
	\subfloat[Stain]{
		\label{fig:stain}
		\includegraphics[width=0.45\textwidth]{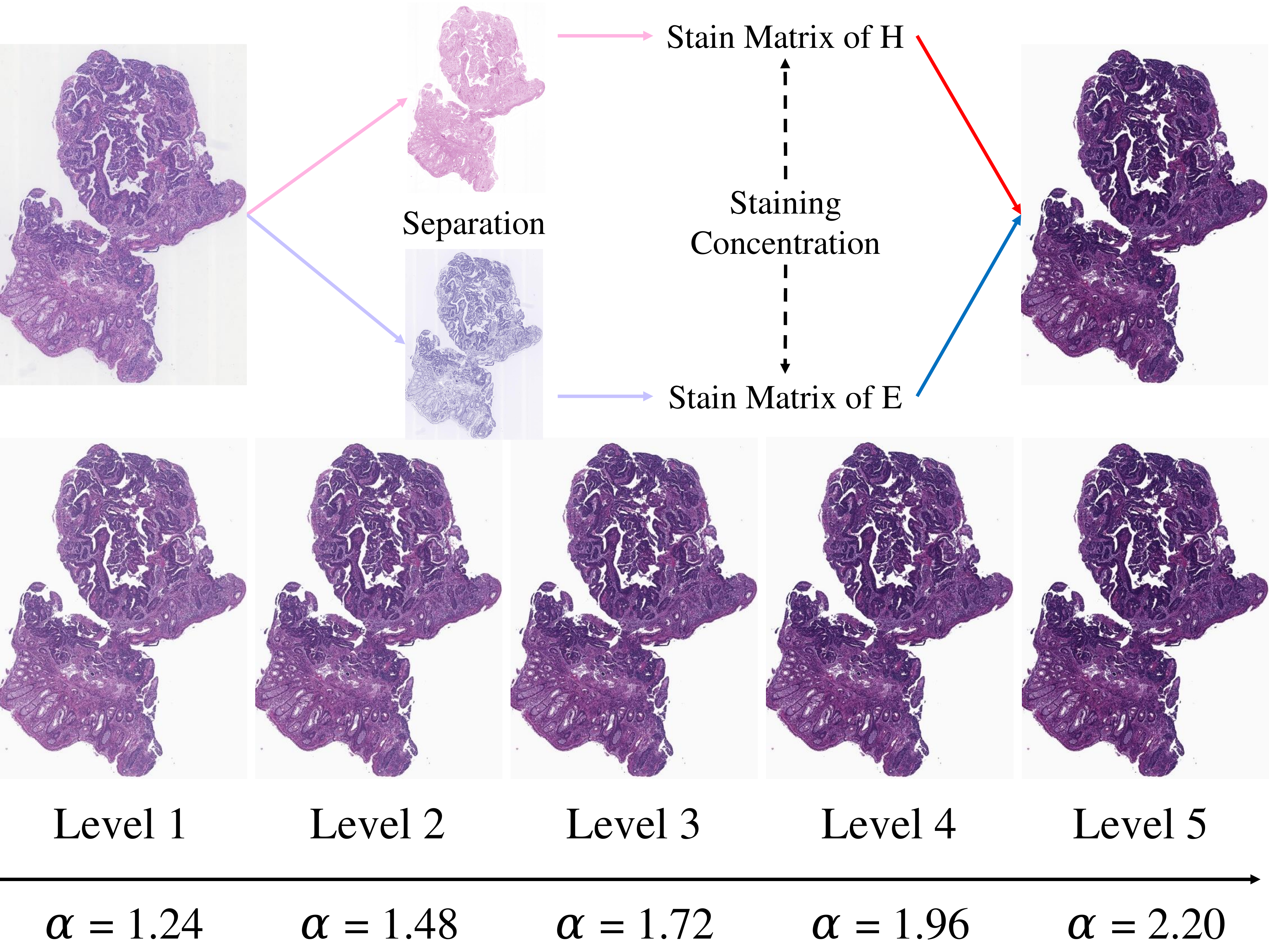}
	}
	\subfloat[Deformation]{
		\label{fig:deformation}
		\includegraphics[width=0.45\textwidth]{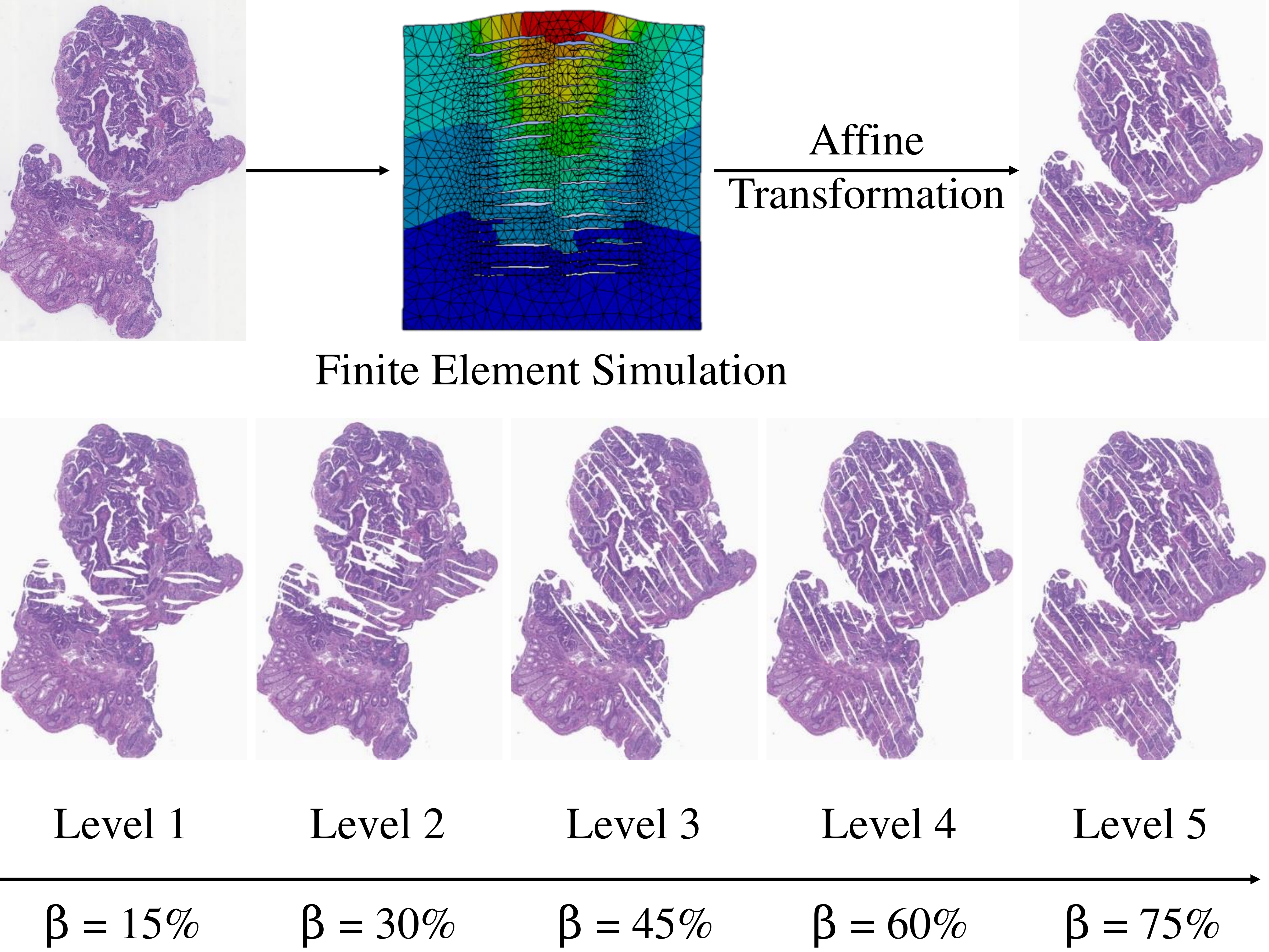}
	}
	\caption{Corruption emulations of 5-level severity. Take \emph{Over-stained with H}\&\emph{E} corruption and \emph{Venetian} corruption for example. $\alpha$ is used for controlling staining concentration and $\beta$ denotes the area scale covered by venetian.}
	\label{fig:simulation example}
\end{figure*}

\section{Omni-Corruption Emulation}
\subsection{Physical Causes of Corruptions}
As shown in Fig. \ref{fig:taskflow}, a full pathological lifecycle contains two main phases: tissue preparation and slide
imaging, each consists of several sub-steps that involve corruption generation from human or device variance, as analyzed below.
\subsubsection{Corruptions in Tissue Preparation}
include \emph{Crack}, \emph{Venetian}, \emph{Knife Line}, \emph{Thick and Thin Section}, \emph{Fold}, \emph{Bubble}, \emph{Stain Deposit}, \emph{Stain Variance}, \emph{Residue}. During sectioning, a defective blade may cause \emph{Crack} (tearing of the tissue structure) or many fine lines called \emph{Knife Lines}. And when disposable blades are not properly supported in the knife holder, tiny vibrations in the knife edge will bring fine parallel cracking like \emph{Venetian}. Besides, a loosely attached knife may form \emph{Thick and Thin Sections} presenting banded areas of different staining levels adjacent to each other. In addition, there are some common impurities, such as the \emph{Stain Deposit} comes with incomplete dissolution of the stains. And because of the unflattened slide, some small \emph{Bubbles} and \emph{Folds} are often be seen. When it comes to the staining, there may be irregular staining during the staining process, such as a change in the concentration or staining time of hematoxylin or eosin, then leads to \emph{Over-staining} or \emph{Under-staining}. Besides, some \emph{Residues} can also lead to locally uneven staining. Sometimes slide dewaxing is incomplete, then \emph{Residual Wax} results in unevenly H\&E stained areas. And inefficient washing after “blueing” will leave \emph{Residual Alkali} resulting in uneven eosin staining. \emph{Residual Xylene} appears when hematoxylin solution rapidly, which causes uneven hematoxylin staining.

\subsubsection{Corruptions in Slide Imaging}
include \emph{Color Cast}, \emph{Exposure}, \emph{Defocus}. \emph{Color Cast} includes \emph{Cold Color} and \emph{Warm Color} caused by the color temperature of the microscope illumination. Besides, uneven illumination will introduce \emph{Underexposure} or \emph{Overexposure}. And \emph{Defocus} is a blur phenomenon due to inaccurate focusing.
	


\subsection{Emulation of Corruptions}
For emulation, we classify the lifecycle corruptions according to three types of their physical causes and the corresponding engines : {Stain} (\emph{Stain Variance}, \emph{Residue}, \emph{Thick and Thin Section}) generated by \textbf{Chemical Engine}, {Deformation}  (\emph{Crack}, \emph{Venetian}, \emph{Fold}) generated by \textbf{Mechanical Engine}, {Color} (\emph{Exposure}, \emph{Color Cast}), {Defocus} and {Coverage} (\emph{Stain Deposit}, \emph{Bubble}, \emph{Knife Line}) generated by \textbf{Optical Engine}.

All of these corruptions are emulated in five levels. And indicators for each level have been discussed and confirmed with different senior pathologists to make ensure that OmniCE-corrupted images within a certain level range are possible in realistic clinical scenarios. After confirming the most severe level, even parameter intervals are assigned to different levels (Fig. \ref{fig:simulation example}).


The Stain corruption refers to the uneven staining concentrations of different stains, or even serious deviation from normal staining. The emulation uses Macenko's method \cite{5193250} to separate the staining concentration of the pathological image and the color matrices of the two stains as shown in Fig. \ref{fig:simulation example}. Then the color matrices of the two stains are multiplied by different coefficients or assign different coefficients to different random areas to vary the staining concentration of different stains in the local area. 

The Deformation corruption involves a large area of the slide. Specially to emulate \emph{Crack} and \emph{Venetian} more realistically, we use ANSYS WORKBENCH \cite{lee2018finite} to emulate the mechanical properties of polymer materials to approximate the deformation of tissue sections when being stretched by various external forces. Firstly, irregular cracks are preset on the model of the thin section. 
After setting the material and Young's modulus on the software, we emulate the applied force and obtain the templates before and after deformation. Then templates are randomly rotated and used for affine transformation of image tiles within each mesh on the templates to get the whole slide image (WSI) after the deformation corruption as shown in Fig. \ref{fig:simulation example}. \emph{Fold} is also achieved by affine transformation of some preset folding templates. The overlapping regions uses sum-up in the optical density space to get a more reasonable visual effect.

For Color corruptions, we adjust the proportion of red and blue channels for \emph{Color Cast} or scale pixel values for \emph{Exposure}. Then \emph{Defocus} corruption is emulated by setting parameters, i.e. RGB illumination wavelength, refractive index and objective NA, to quantitatively generate point spread functions (PSF) of RGB center wavelengths at different locations, and then convolving images of corresponding color channels with defocus PSFs. And for Coverage corruptions, we overlay with preset templates in the optical density space.

Due to the limited number of pages, we will organize and open-source all the detailed formulas, related quantitative parameters and corruptions generating codes involved here after publication as soon as possible.


\begin{table*}[t]
    \centering
    \caption{Error rate results ($\downarrow$) of the patch level OmniCE corruption benchmark datasets. 
    The best and worst results among models are marked by \textbf{value} and \underline{value} for every row, \textcolor{blue}{value} and \textcolor{red}{value} for every column.}
    \resizebox{2\columnwidth}{!}{
    \tablestyle{1pt}{1}
    \begin{tabular}{l|rr|rr|rr|rr|rr|rr||rr|rr|rr|rr|rr|rr}
    \toprule\toprule
     \multicolumn{1}{c}{}&\multicolumn{12}{c||}{PatchCamelyon} &\multicolumn{12}{c}{CRC-HE}\\
    \toprule\toprule
    
         \multirow{2}{*}{OmniCE}& \multicolumn{2}{c|}{AlexNet}& \multicolumn{2}{c|}{VGG16} & \multicolumn{2}{c|}{ResNet18} & \multicolumn{2}{c|}{ResNet50} & \multicolumn{2}{c|}{ResNet101} & \multicolumn{2}{c||}{DenseNet121}& \multicolumn{2}{c|}{AlexNet} & \multicolumn{2}{c|}{VGG16} & \multicolumn{2}{c|}{ResNet18} & \multicolumn{2}{c|}{ResNet50} & \multicolumn{2}{c|}{ResNet101} & \multicolumn{2}{c}{DenseNet121} \\ 
       &mCE&rCE&mCE&rCE&mCE&rCE&mCE&rCE&mCE&rCE&mCE&rCE&mCE&rCE&mCE&rCE&mCE&rCE&mCE&rCE&mCE&rCE&mCE&rCE\\
        \midrule\midrule
        
        Under-stained  H\&E & \underline{100.0} & \textbf{2.3} & 55.8 & 3.9 & \textbf{45.3} & 3.1 & 55.6& 3.7 & 55.1 & 3.7 & 62.6 & \underline{5.1} & 100.0 & 4.6& \textbf{48.0} & 4.8 & 75.4 & 4.6 & 71.1& \textbf{4.1} &\underline{111.6}  &\underline{7.4}  & 67.1&5.3 \\ 
        Over-stained  H\&E& \underline{100.0} & 2.0  & \textbf{31.3} & \textbf{1.9} & 37.6 & 2.3 & 55.6& \underline{3.3} & 44.1 & 2.6 & 46.1 & \underline{3.3}& 100.0 & \textbf{3.3}& \textbf{54.6} & 3.9 & \underline{140.7}& 6.1 & 104.5 &4.4 & 114.8 &5.5  & 113.4 &\underline{6.4}  \\  
         Under-stained  H& \underline{100.0} & \textbf{2.3}   & 59.7 & 4.2 & \textbf{52.9} & 3.7 & 57.6& 3.8 & 53.4 & 3.6 & 53.6&\underline{4.4} & 100.0 & \textbf{4.6}& \textbf{81.5} & 8.1 & 97.5 & 5.9 & 108.9& 6.3 &\underline{128.4}  &\underline{8.5}  &88.0 &6.8 \\ 
         Over-stained  H& \underline{100.0} & 1.9   & 32.0 & 1.8 & 36.6 & 2.1 & 44.1& \underline{2.4} & 28.9 & \textbf{1.6} & \textbf{27.3} & 1.8& 100.0 & \textbf{2.7}& \textbf{64.9} & 3.8 & 141.0& 5.1 & 151.9 &5.2 & \underline{154.3} &\underline{6.1}  & 102.7 &4.8  \\  
         Under-stained  E& \underline{100.0} & \textbf{1.5}   & 42.2 & 1.9 & \textbf{34.2} & 1.6 & 39.6& 1.7 & 35.0 & \textbf{1.5} & 36.8&\underline{2.0}& \underline{100.0} & \underline{2.4}& 47.4 & \underline{2.4} &66.7 &2.1  & 54.8 &\textbf{1.6} &49.7  &1.7  & \textbf{44.3} &1.8  \\  
         Over-stained  E& \underline{100.0} & \textbf{2.3}   & \textbf{43.4} & 3.0 & 61.3 & 4.2 & 78.6& 5.2 & 51.3 & 3.4 & 66.4 & \underline{5.4}& 100.0 & \textbf{3.2}& \textbf{67.6} & 4.7 &\underline{151.5} &6.4  &122.8  &5.0 &  110.3&5.1  & 126.5 &\underline{6.9}  \\  
         Residual Wax& \underline{100.0} & \textbf{1.7}   & 37.4 & 1.9 & \textbf{33.1} & \textbf{1.7} & 37.2& 1.8 & 36.2 & 1.8 & 40.6 & \underline{2.4}& \underline{100.0} & 1.6 & \textbf{37.4} & 1.3 &58.0 &\textbf{1.2}  &67.0  &1.4 & 94.4 &\underline{2.2}  &53.3  & 1.5 \\ 
         Residual Xylene& \underline{100.0} & 1.7   & 34.2 & 1.8 & \textbf{30.7} & \textbf{1.6} & 35.6& 1.7 & 35.9 & 1.8 & 31.1 & \underline{1.9}& 100.0 & 1.5 & \textbf{38.5} & 1.2 &58.7 &\textbf{1.1}  &73.4  &1.4 & \underline{113.8} &\underline{2.4}  &68.4  &1.7  \\  
         Residual Alkali& \underline{100.0} & \textbf{1.3}   & 36.3 & 1.4 & 32.8 & \textbf{1.3} & 37.8& 1.4 & 36.1 & \textbf{1.3} & \textbf{32.3} & \underline{1.5}& \underline{100.0} & \underline{1.3} & \textbf{43.2} & 1.2 & 56.2&\textbf{\textcolor{blue}{1.0}}  & 62.0 &\textbf{\textcolor{blue}{1.0}} & 60.2 &1.1  &50.1  &1.1  \\ 
         Thick and Thin& \underline{100.0} & \textbf{1.8}   & 35.2 & 1.9 & \textbf{33.9} & \textbf{1.8} & 42.3& 2.1 & 39.5 & 2.0 & 42.8 & \underline{2.7}& 100.0 & 2.4 & \textbf{39.4} & \textbf{2.0} &105.0 &3.3  &86.3  &2.6 &\underline{112.7}  &\underline{3.9}  & 91.0 &3.7  \\  
        \midrule
        Stain Deposit& \underline{100.0} & \underline{1.4}   & \textcolor{blue}{25.6} & \textcolor{blue}{1.1} & \textcolor{blue}{23.5} & \textbf{\textcolor{blue}{1.0}} & \textcolor{blue}{26.9}& \textcolor{blue}{1.1} & \textcolor{blue}{26.1} & 1.1 & \textbf{\textcolor{blue}{22.5}} & \textcolor{blue}{1.1} & \underline{100.0} & \underline{1.2} & \textbf{42.8} & 1.1 & 65.1 & \textbf{\textcolor{blue}{1.0}} &75.2 &\underline{1.2}  & 61.7 &1.1  & 53.3&1.1 \\ 
         Bubble& \underline{100.0} & \textbf{\textcolor{blue}{1.1}}   & 42.9 & 1.4 & 46.5 & 1.5 & 53.7& \underline{1.6} & 51.0 & \underline{1.6} & \textbf{37.4} & 1.4& \underline{100.0} & \textcolor{blue}{1.1} & 66.2 & \underline{1.6} &78.5 &1.2  &71.2  & \textbf{\textcolor{blue}{1.0}}& 64.3 &\textbf{\textcolor{blue}{1.0}}  & \textbf{63.4} &1.2  \\  
        
         Knife Line& \underline{100.0} & \underline{\textcolor{blue}{1.1}}   & 35.5 & \underline{\textcolor{blue}{1.1}} & 31.8 & \textbf{\textcolor{blue}{1.0}} & 34.6& \underline{\textcolor{blue}{1.1}} & 33.6 & \textbf{\textcolor{blue}{1.0}} & \textbf{29.1} & \underline{\textcolor{blue}{1.1}}& \underline{100.0} & \underline{1.5} & \textbf{\textcolor{blue}{31.5}} & \textbf{\textcolor{blue}{1.0}} & 49.7&\textbf{\textcolor{blue}{1.0}}  & 54.8 &1.1 & \textcolor{blue}{44.9} &\textbf{\textcolor{blue}{1.0}}  & 39.9 &\textbf{\textcolor{blue}{1.0}}  \\  
        Crack& \underline{100.0} & \textbf{1.2}  & 53.6 & 2.0 & 56.3 & 2.1 & 54.5& 1.9 & 50.7 & 1.8 & \textbf{49.3} & \underline{2.2} & \underline{100.0} & \underline{2.5} & 40.2 & 2.2 & \textcolor{blue}{38.5} &\textbf{1.3}  & \textcolor{blue}{49.0}& 1.5 & 50.0 &1.8  &\textbf{\textcolor{blue}{32.7}} &1.4 \\ 
        \midrule\midrule
        
         Cold Color& 100.0 & \textbf{2.0}   & 88.6 & 5.3 & \textbf{71.3} & 4.3 & 84.6& 4.8 & \underline{106.3} & \underline{6.1} & 79.0 & 5.6& 100.0 & 3.2 & \textbf{43.6} & \textbf{3.1} & 102.9&4.4  & 113.4 &4.6 & 153.0  & 7.1 & \underline{191.8} &\underline{10.6}  \\  
         Warm Color& \underline{100.0} & \textbf{\textcolor{red}{3.0}}   & \textbf{83.1} & 7.6 & 89.9 & \textcolor{red}{8.2} & 91.0& \textcolor{red}{7.9} & 85.5 & 7.5 & 84.5 & \underline{\textcolor{red}{9.1}} & \underline{100.0} & \textbf{\textcolor{red}{7.5}} & \textbf{77.8} & \underline{\textcolor{red}{12.6}} & 93.6 &9.2  & 93.2& \textcolor{red}{8.8} & 95.5 & 10.3 & 82.6 &10.5 \\ 
         Overexposure& \textbf{100.0} & \textbf{2.0}   & \textcolor{red}{127.3} & \textcolor{red}{7.7} & 112.7 & 6.8 & \textcolor{red}{125.9}& 7.2 & \underline{\textcolor{red}{142.2}} & \textcolor{red}{8.2} & \textcolor{red}{121.1} & \underline{8.6} & \textbf{100.0} & \textbf{3.0} & \textcolor{red}{168.9} & 11.0 &\textcolor{red}{251.9}  &\textcolor{red}{9.9}  &\textcolor{red}{226.9} &8.6  & \underline{\textcolor{red}{280.1}} &\textcolor{red}{12.1}  & \textcolor{red}{263.5}&\underline{\textcolor{red}{13.5}} \\ 
         Underexposure& 100.0 & \textbf{1.7}  & \textbf{90.9} & 4.8 & \textcolor{red}{117.3} & 6.2 & 114.3& 5.7 & \underline{136.1} & \underline{6.9} & 107.3 & 6.6 & 100.0 & \textbf{2.0} & \textbf{78.8} & 3.4 & 79.4 &2.1  & \underline{154.6}& \underline{3.9} & 124.8 &3.6  &94.5 &3.2 \\

        Defocus& 100.0 & \textbf{2.4}   & 97.0 & 7.0 & 97.4 & 7.0 & \underline{100.6}& 6.9 & \textbf{81.7} & 5.7 & 90.4 & \underline{7.7}& \textbf{100.0} & \textbf{2.7} & 100.9 & 5.9 &150.8 &5.4  & 128.3 &4.4 & \underline{206.7} &\underline{8.1}  &164.2  & 7.6 \\  
         \midrule\midrule
        Average& \underline{100.0} & \textbf{1.8}   & 55.4 & 3.3 & \textbf{55.0} & 3.2 & 61.6& 3.4 & 59.4 & 3.3  & 55.8 & \underline{3.9}& 100.0 & \textbf{2.8} & \textbf{61.7} & 4.0 & 98.0&3.8 & 98.4& 3.6 & \underline{112.2} & \underline{4.7} & 94.3 & \underline{4.7}\\ 
        Original Error & \multicolumn{2}{c|}{\underline{12.9}} & \multicolumn{2}{c|}{4.2} & \multicolumn{2}{c|}{4.2} & \multicolumn{2}{c|}{4.5}& \multicolumn{2}{c|}{4.4}&  \multicolumn{2}{c||}{\textbf{3.6}} &\multicolumn{2}{c|}{\underline{8.4}}& \multicolumn{2}{c|}{\textbf{3.9}} & \multicolumn{2}{c|}{6.4} & \multicolumn{2}{c|}{6.7}& \multicolumn{2}{c|}{5.9}& \multicolumn{2}{c}{5.0} \\ 
        
        \bottomrule\bottomrule
    \end{tabular}
    }
    \label{table:pcam2}
\end{table*}

\section{Experiment}

\subsection{OmniCE Corruption Benchmark Datasets}
We collect two different types of pathology datasets, which are patch level and slide level, for omni-corruption emulation and robustness evaluation. Then we train on clean images and test on these benchmark datasets with corruption. 


\subsubsection{Patch-level Dataset}
Given consideration to different tissue types, we select two datasets derived from lymph node and colon tissue respectively, i.e. Patchcamelyon \cite{Veeling2018-qh} and CRC-HE dataset \cite{kather_jakob_nikolas_2018_1214456}, for benchmarking.

In PatchCamelyon, patches with the size of 96$\times$96 are extracted from slides of potentially metastatic breast cancer, the label is whether the patch contains tumor.
Here, we have removed duplicates and normalize the staining style of images from different centers (1,2,3) for more rational grading of staining corruptions. Finally, we obtain 208,401 training examples and 10,000 remaining examples for omni-corruption emulation as synthetic distribution shift to benchmark the robustness. The data from Center 4, 5 are used to benchmark the natural distribution shift. 
The CRC-HE dataset contains 100,000 patches with 224$\times$224 pixels divided into 9 classes for training.
And for the validation, there are 7180 patches with colorectal adenocarcinoma which are used to emulate 19 different types of corruptions.

\begin{figure*}[!t]
	\centering
	\subfloat[PatchCamelyon]{
		\label{fig:pcam-drop}
		\includegraphics[width=0.3\textwidth]{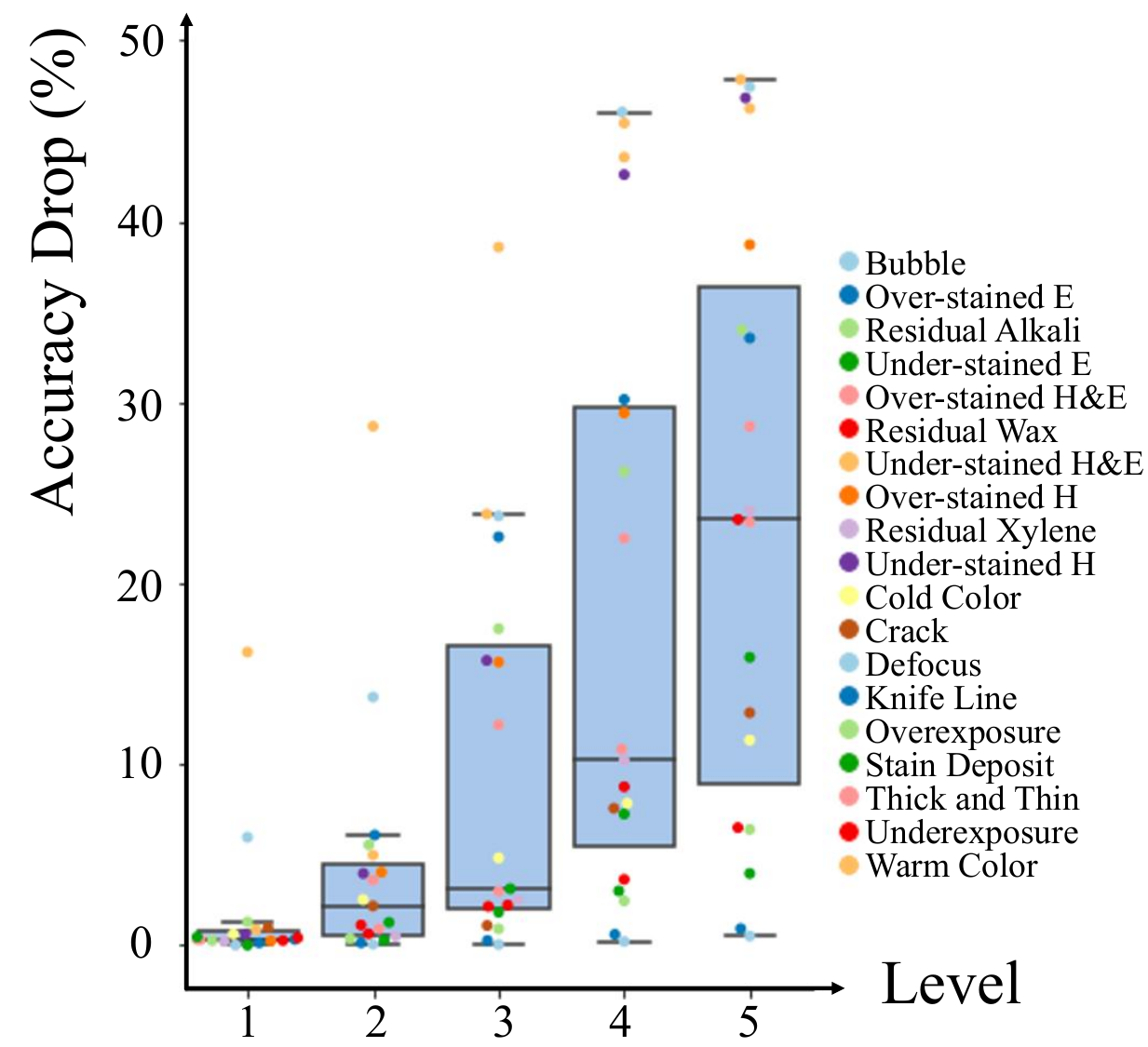}
	}
	\subfloat[CRC-HE]{
		\label{fig:crc-drop}
		\includegraphics[width=0.3\textwidth]{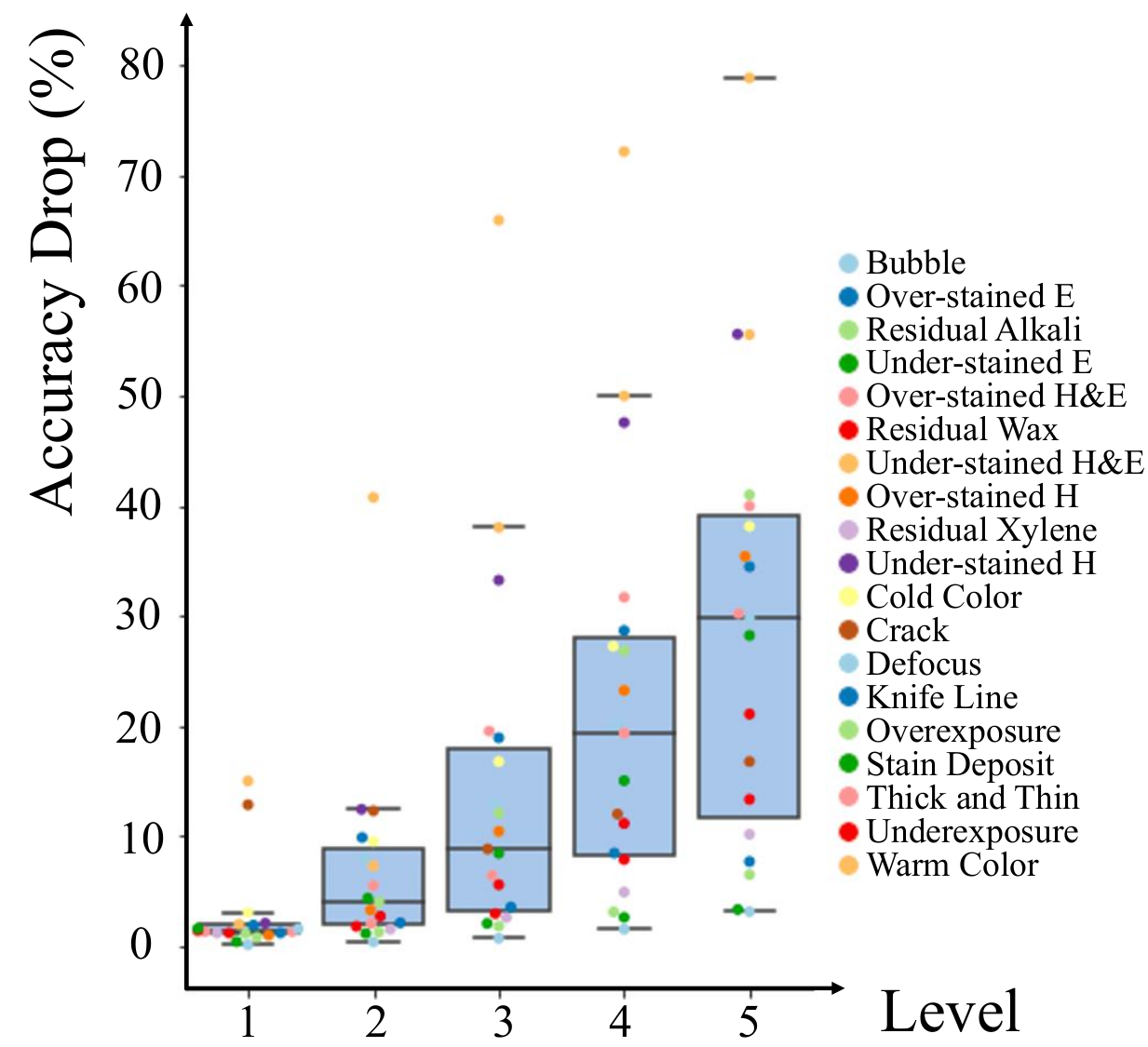}
	}
        \subfloat[DigestPath]{
		\label{fig:seg_drop}
		\includegraphics[width=0.3\textwidth]{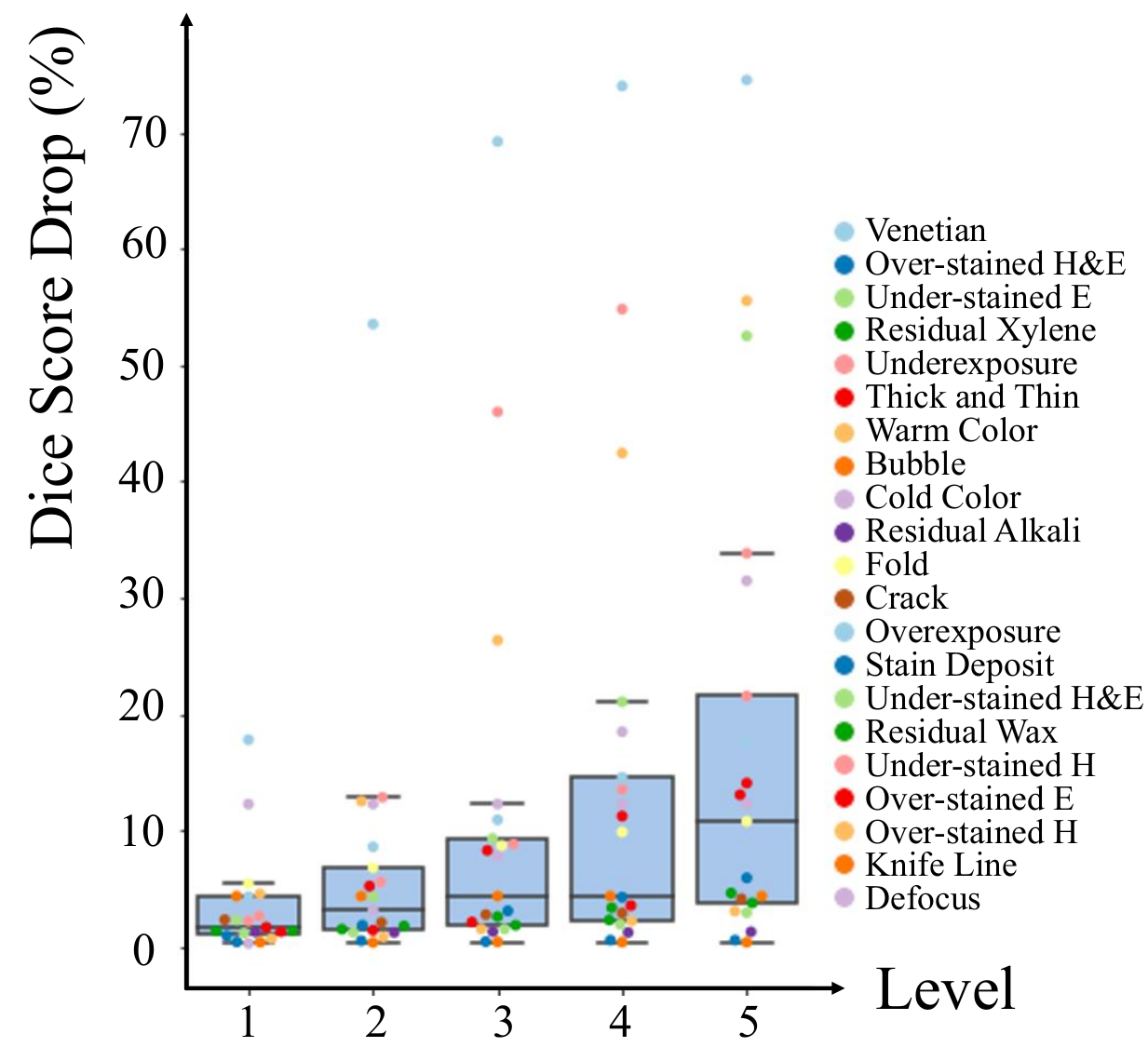}
	}
	\caption{Box plots of dropped metric values with different severity levels on different benchmarks. Dropped metric values are calculated by subtracting metric values of corrupted images from metric values of clean images.}
	\label{fig:acc_drop}
\end{figure*}
We use AlexNet \cite{krizhevsky2017imagenet}, VGG16 \cite{simonyan2014very}, ResNet18/50/101 \cite{he2016deep} and DenseNet121 \cite{huang2017densely} for training and testing. As for metrics, error rate $Error^{f}$ of each model $f$ is computed for original images and $CE_{s,c}^{f}$ for each corruption type $c$ and severity level $s$, we can then get the mean corruption error rate $mCE_{c}^{f}$ for a single corruption type $c$ by taking the average across all severity levels for that corruption. As presumably not all corruptions are equally difficult, we adjust by a baseline which in our case is AlexNet’s corruption error rate $CE_{s, c}^{AlexNet}$. Thus, we get:
\begin{equation}
    mCE_{c}^{f} = (\sum_{s=1}^{5} CE_{s, c}^{f})/(\sum_{s=1}^{5} CE_{s, c}^{AlexNet})
\end{equation}
And we also introduce:
\begin{equation}
    rCE_{c}^{f} = (\frac{1}{5}\sum_{s=1}^{5} CE_{s, c}^{f})/(Error^{f})
\end{equation}
It measures how gracefully a classifier degrades in the presence of corruptions. 


 \subsubsection{Slide-level Dataset}
The DigestPath dataset \cite{d9aabcc6baf0458996bfb65dfb3e5bb7} is used for slide level colorectal tissue segmentation, including 872 tissue sections with an average size of 3000 $\times$ 3000. We take 172 slide images color-normalized globally for OmniCE. In addition, two extra corruptions (\emph{Fold} and \emph{Venetian}) are emulated specific to slide images. The two corruptions are also common in clinical settings, they are more suitable for application and quantification on slide images and has great significance in studying the impact of corruptions on segmentation. It is worth noting that both original images and masks are deformed in the same way.
The following segmentation models, UNet \cite{ronneberger2015u}, Deep Contour-aware Network (DCAN) \cite{chen2016dcan}, Global Convolutional Network (GCN) \cite{peng2017large} and Dense-UNet \cite{cai2020dense} are chosen to train and test, and the dice score is used as the main metric. 




\subsection{Robustness Evaluation}

\subsubsection{Patch-level Benchmark}
As shown in \Cref{table:pcam2}, our experiment investigates effects of corruptions on model performance in the patch classification task. 
For corruptions in the top six rows of the table, which emulated by the staining engine, we observe that under-stained with H\&E patches result in lower model performance compared to over-stained with H\&E patches, since what is inline with our common sense, the under-stained patches may reduce contrast information that is critical for discrimination. We also find that over-stained with H patches result in higher model performance compared to under-stained with H patches, which is expected given more distinguishable nuclei. Conversely, over staining with E will indirectly affect the contrast of nuclei, which has a negative impact on model performance.
For other 4 stain-related corruptions, both the mCE and rCE scores are lower than the under-stained ones due to smaller stain-corrupted regions, while the staining shift of the whole patch severely degrades model discrimination. For the next 4 deformation and coverage corruptions, we find that they have a more subtle effect on model performance compared to stain-related corruptions. 
And the deformation corruption has a relatively strong effect on model performance compared to coverage corruptions due to the fact that the deformation engine can change the shape of some areas in the image, while coverage corruptions mainly result in the loss of some pixels. 

 \begin{table}[t]
    \centering
    \tiny
    \caption{Dice score results ($\uparrow$) of the slide level OmniCE corruption benchmark dataset. 
    The best and worst results are marked in the same way as in Table \ref{table:pcam2}.}
    \resizebox{0.99\columnwidth}{!}{
    \tablestyle{10pt}{1.0}
    \begin{tabular}{@{}l|cccc}
    \toprule\toprule
        OmniCE & UNet & GCN & DCAN & Dense-UNet \\ %
        \midrule\midrule
        
        Under-stained H\&E & \textbf{0.6138} & 0.5668 & \underline{0.3968} & 0.5724  \\ 
         Over-stained  H\&E & \textbf{0.7438} & 0.7137 & \underline{0.6088} & 0.7194 \\ 
         Under-stained H & \textcolor{red}{0.3450} & 0.4454 & \underline{0.2415} & \textbf{0.4496}  \\ 
         Over-stained  H & \textbf{0.7435} & 0.7291 & \underline{0.6378} & 0.7416  \\ 
         Under-stained E & \textbf{0.7620} & 0.7283 & \underline{0.6183} & 0.7349  \\ 
         Over-stained  E & 0.6312 & 0.6649 & \underline{0.4729} & \textbf{0.6768}  \\ 
         Residual Wax & \textbf{0.7690} & 0.7243 & \underline{0.6560} & 0.6950  \\
         Residual Xylene & \textbf{0.7620} & 0.7184 & \underline{0.6482} & 0.6882 \\ 
         Residual Alkali & \textbf{0.7736} & 0.7332 & \underline{0.6635} & 0.7404 \\
         Thick and Thin & \textbf{0.7535} & 0.7030 & \underline{0.6281} & 0.7091 \\ 
         \midrule
         Stain Deposit & \textbf{0.7749} & \underline{0.7408} & \textbf{0.7749} & 0.7504  \\ 
         Bubble & \textcolor{blue}{0.7750} & \underline{\textcolor{blue}{0.7419}} & \textbf{\textcolor{blue}{0.7751}} & \textcolor{blue}{0.7539} \\ 
        
         Knife Line & 0.6912 & \textbf{0.7023} & \underline{0.6721} & 0.7015 \\ 
         Crack & \textbf{0.7651} & 0.7174 & \underline{0.6690} & 0.7110  \\ 
         Fold & \textbf{0.7253} & 0.6626 & \underline{0.6254} & 0.6499\\ 
         Venetian & \textbf{0.6813} & 0.6340& \underline{0.6183} & 0.6512\\ 
         \midrule\midrule

         Cold Color & \textbf{0.7727} & 0.6234 & \underline{0.2383} & 0.6337 \\ 
         Warm Color & \textbf{0.6745} & 0.4629 & \underline{0.1168} & \textcolor{red}{0.3573}\\ 
         Overexposure & \textbf{0.3937} & \textcolor{red}{0.1673}& \underline{\textcolor{red}{0.1100}}& 0.3614 \\ 
         Underexposure & \textbf{0.6622} &0.6424 & \underline{0.3888} & 0.6098 \\
        Defocus & \textbf{0.7181}  & 0.6231  &  \underline{0.612} & 0.6731 \\ 
        \midrule\midrule
        Average & \textbf{0.6920} & 0.6402 &  \underline{0.5320}& 0.6467 \\ 
       Original & 0.7812 & \underline{0.7457} & \textbf{0.7831} & 0.7583 \\
        
        \bottomrule\bottomrule
    \end{tabular}
    }
    \label{table:seg}
\end{table}
We also explore the effects of optical imaging corruptions and our results indicate that these corruptions have the greatest impact on model performance. The color temperature degradation introduces the domain shift, which has more devastating effects on models that are trained and fitted better on the original image domain, such as DenseNet121. Furthermore, overexposure and defocus corruptions are found to be more detrimental to model performance, likely due to the greater amount of information loss. 

Regarding the above conclusion, it can also be seen more intuitively from Fig. \ref{fig:acc_drop}.



\subsubsection{Slide-level Benchmark}
As for the segmentation benchmark shown in \Cref{table:seg}, the trend of most corruptions on model performance is consistent with the previous analysis of \Cref{table:pcam2}.
But for coverage corruptions, they may block the local features for discriminating the image category at the patch level. But at the slide level, semantic segmentation tends to capture the global feature for structural discrimination, hence, the local coverage will cause a lower impact on the segmentation task. By the same token, deformation corruptions disrupt the original morphology of the tissue slice and bring more serious impact on the global feature extraction than individual patches.
\subsubsection{Corruption Level and Performance Drop }
What's more, we explore the model's robustness to image features under different levels of corruption in Fig. \ref{fig:acc_drop}. For different benchmark datasets, the performance drop of the model shows a strong consistency with the level of corruption, that is, images with higher corruption levels are more difficult to be classified by the model. 

In summary, performances of models on these corrupted images conform to common sense and pathological prior knowledge, proving that our OmniCE can effectively emulate corruptions of different causes in real scenes, and the level of corruption accurately reflects the quality of image features.


\subsection{Augmentation with OmniCE-Corrupted Dataset}
In addition, we have explored data augmentation in conjunction with our corruptions and the training dataset of PatchCamelyon \cite{Veeling2018-qh}, which is obtained from three hospital centers, is chosen for model training, then the data from two other different centers is used for testing.
We train with SE-ResNeXt101 \cite{hu2018squeeze} without pretrained models as our baseline and then replace original corruptions on natural images with OmniCE corruptions proposed in this paper for the operation pool of Augmix \cite{hendrycks2019augmix}, which effectively applies corruptions for data augmentation, to leverage support of the OmniCE-corrupted dataset. 

As shown in Table \ref{table:aug}, our method further enhances the model generalization compared to common Augmix and surpasses the existing methods including different augmentations (e.g., normal H\&E jitter) on the leaderboard.


\begin{table}[t]
    \caption{Performance comparison on Camelyon val set and test set.}
    \centering
    \resizebox{0.99\columnwidth}{!}{
    \tablestyle{1.0pt}{1.0}
    \begin{tabular}{@{}llcc@{}}
        \toprule
        Algorithm & Backbone & Val Acc(Center 4)& Test Acc (Center 5) \\ \midrule
        CORAL \cite{koh2021wilds}             & DenseNet121   & 86.2 & 59.5 \\
        IRM \cite{koh2021wilds}              & DenseNet121   & 86.2 & 64.2 \\
        CGD \cite{piratla2021focus}        & DenseNet121   & 86.8 & 69.4 \\
        Fish \cite{shi2021gradient}             & DenseNet121   & 83.9 & 74.1 \\
        LISA \cite{yao2022improving}             & DenseNet121   & 81.8 & 77.1 \\
        ERM w/ data aug \cite{sagawa2021extending}   & DenseNet121  & 90.6 & 82.0 \\
        ERM w/ targeted aug \cite{gao2022out}   & DenseNet121  & 92.7 & 92.1 \\
        ERM w/ H\&E jitter \cite{8327641} & SE-ResNeXt101  & 88.0 & 91.6 \\
        \midrule
        Normal Training                & SE-ResNeXt101 & 75.3     & 57.2     \\
        Augmix \cite{hendrycks2019augmix}            & SE-ResNeXt101 & 86.6     & 76.9     \\
        \textbf{Ours (+OmniCE)} & \textbf{SE-ResNeXt101} & \textbf{94.9}     & \textbf{92.2}     \\ 
        \bottomrule
        \end{tabular}
    }
    \label{table:aug}
\end{table}

\section{Conclusion}
In this paper, we firstly analyse physical causes of 21 types of
corruptions throughout the pathological life-cycle and propose an OmniCE emulator to construct benchmark datasets for evaluating the robustness of typical DNNs in digital pathology. Furthermore, the OmniCE-corrupted dataset is used for data augmentation during model training, which is validated on the multicenter data of Camelyon and obtains a significant improvement in generalisation capability.

\section*{Acknowledgment}
We thank Qiuchuan Liang (Beijing Haidian Kaiwen Academy, Beijing, China) for preprocessing data.






{
\bibliographystyle{IEEEtran}
\bibliography{IEEEabrv}
}


\end{document}